\documentclass[%
 reprint,floatfix,
 amsmath,amssymb,
 aps,inkscapelatex=false,
]{revtex4-2}

\usepackage{graphicx}
\usepackage{dcolumn}
\usepackage{bm}
\usepackage{svg}
\usepackage{enumerate}
\usepackage{xcolor}
\usepackage{comment}
\usepackage{svg}
\usepackage{hyperref}
\usepackage{comment}

\begin{document}

\title{Bifurcations of inflating balloons and interacting hysterons}

\author{Gentian Muhaxheri}
\email{gmuhaxhe@syr.edu}

\author{Christian D. Santangelo}%

\affiliation{%
Department of Physics,
Syracuse University, Syracuse, New York, 13244
}%


\begin{abstract}
While many materials exhibit a complex, hysteretic response to external driving, there has been a surge of interest in how the complex dynamics of internal materials states can be understood and designed to process and store information. We consider a system of connected rubber balloons that can be described by a Preisach model of non-interacting hysterons under pressure control, but for which the hysterons become coupled under volume control. We study this system by exploring the possible transition graphs, as well as by introducing a configuration space approach which tracks the volumes of each balloon. Changes in the transition graphs turn out to be related to changes in the topology of the configuration space of the balloons, providing a particularly geometric view of how transition graphs can be designed, as well as additional information on the existence of hidden metastable states. This class of systems is more general than just balloons.
\end{abstract}

\maketitle

\section{\label{sec:level1}Introduction}

Many systems in condensed matter, such as amorphous media \cite{memory1, memory2, memory3,keim2020global,adhikari2018memory}, crumpled paper \cite{memory4,matan2002crumpling}, corrugated sheets \cite{corrugated}, multistable origami \cite{origamiswitches,forte2021mua} and mechanical metamaterials \cite{chen2021reprogrammable, memory5,sun2019snap,rafsanjani2016snapping,yang2019multi}, exhibit ``memory'' encoded by the configurations of internal states.
These systems are often modeled as a collection of primordial, bistable elements called hysterons, which switch between two states according to the history of some driving field (Fig.\ref{fig:hysteron}a). The sequence of how individual hysterons switch states gives rise to important collective effects such as return point memory \cite{memory6}.
The Preisach model of non-interacting hysterons is the prototypical model for describing how individual hysterons lead to collective hysteresis \cite{preisach1935magnetische}. When interactions between hysterons are introduced, however, the range of behavior expands dramatically: one sees multiperiodic orbits \cite{memory1, memory2}, scrambling and avalanches \cite{corrugated, van2021profusion}, transient memory \cite{memory5, memory6}, and the ability to mimic finite-state machines \cite{adamczyk2003anthology}.

In this paper, we will consider a particular class of interacting hysteron systems based on inflating rubber balloons. In many types of balloons, the pressure depends non-monotonically on the volume \cite{verron2003numerical}, and consequently balloons at constant pressure can exhibit bistability and be modeled as hysterons. When several balloons are joined in parallel to share a common volume of air, however, the total pressure becomes dependent on the state of the other balloons, introducing an effective global interaction between hysterons. We will show that this interacting hysteron system has an alternate description, rooted in the geometry of a system of curves in a configuration space parameterized by the volumes of the individual balloons. Tuning the response of individual balloons changes the topology of the configuration space curves through a process of bifurcation-mediated recombination. Different topological classes of configuration spaces correspond to different possible transition graphs that can be realized.

While balloons are interesting in their own right, having a number of applications to medicinal surgery \cite{sinuplasty, ciernik2002line, glozman2010self}, automobile air bags, pneumatic actuation and shape morphing \cite{pikul2017stretchable, baromorphs, baines2023programming}, and soft robotics \cite{gorissen2019hardware, chi2022bistable, van2023nonlinear}, here we think of them as a prototype for an entire class of systems which also includes bistable origami \cite{origamiswitches} and mechanical beams \cite{adamczyk2003anthology} under fixed strain.

The common element is that the bistability of the individual hysterons is governed by a smooth energy. In the case of balloons, this means that the pressure is uniquely determined by the volume; for buckled beams this will mean that the force is determined uniquely by the displacement of the central beam.
Our approach provides an intriguing link between two seemingly distinct methods to study the complex behavior of designed materials: transition graphs induced by hysteretic state changes, and the bifurcations of smooth configuration spaces. Indeed, we will explicitly show how to obtain transition graphs using our configuration space approach.

In Sec. \ref{sec:level1}, we briefly discuss non-monotonic inflation in balloons and interpret this through non-interacting hysterons when the balloons are held at constant pressure and globally interacting hysterons when the balloons have a constant, shared volume. In Sec. \ref{sec:configspace} we will define and derive the configuration space for interacting balloons and develop our understanding of how the configuration space topology changes through bifurcations. Finally, in Sec. \ref{sec:connect}, we demonstrate the connection explicitly.

\section{\label{sec:level1}History-dependent behavior of balloons}

It is well-known that in a typical rubber balloon, the pressure depends on the inflated volume of the balloon non-monotonically according to a function $P(V)$ which is, itself, the derivative of an energy \cite{plants, verron2003numerical,levin2004two}. At a critical volume, $V_+$, the pressure obtains a local maximum. Above $V_+$, the pressure decreases until a higher volume, $V_-$, after which it rises again. In party balloons, this is experienced as the balloon becoming easier to inflate after some air has been put into the balloon. Were we to inflate the balloon at constant pressure, it would exhibit a sudden increase in volume at a critical pressure $P(V_+)=P_+$ and, upon deflation, a sudden decrease in volume at a lower pressure, $P(V_-)=P_-$. Thus, a single balloon can exhibit hysteresis and a state can be assigned to each balloon (Fig. \ref{fig:hysteron}).

\subsection{Pressure control}

\begin{figure}[]
  \centering
  \includegraphics[width=8.6cm]{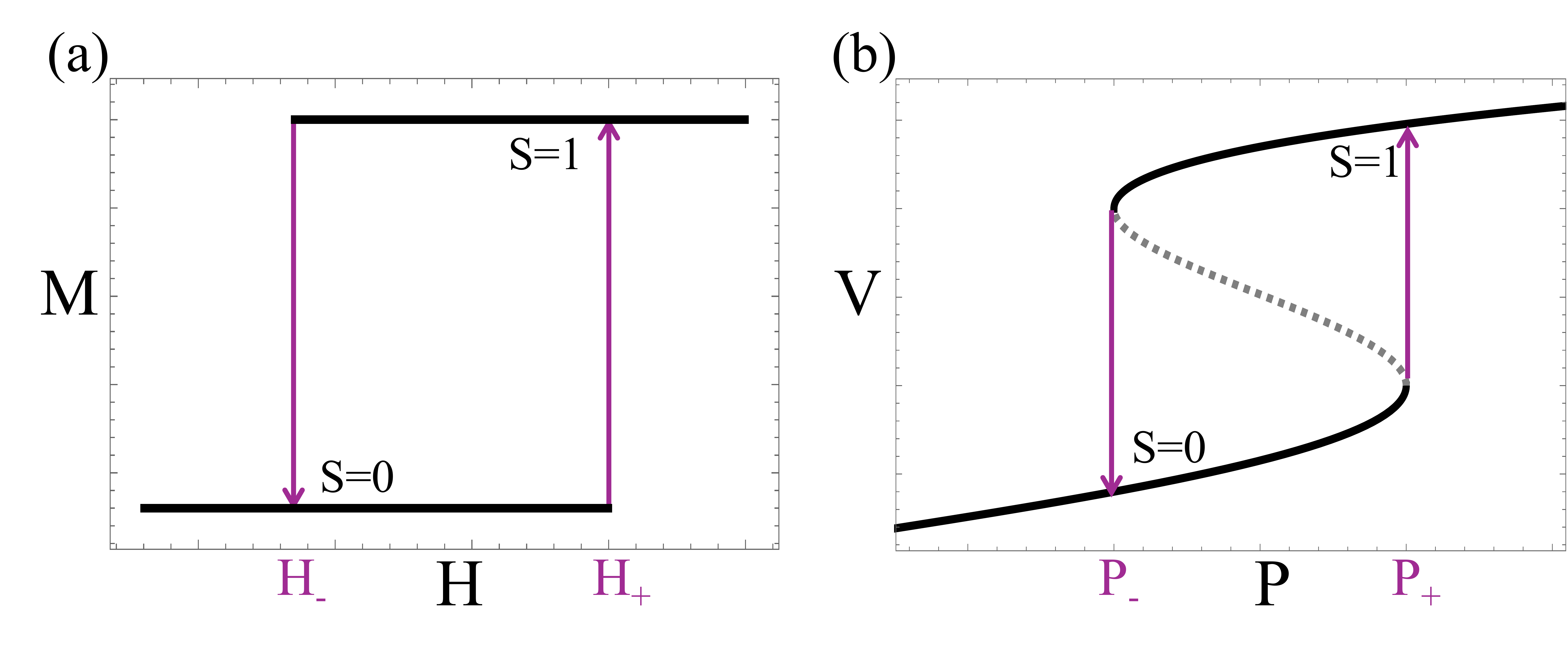}
  \caption{(a) Schematic of a single hysteron showing its two stable configurations, $0$ and $1$. (b) The pressure-volume curve for a rubber balloon showing that there is a jump between two different volumes at given threshold pressure.}
  \label{fig:hysteron}
\end{figure}

When $N$ non-identical balloons are connected to a pump held at constant pressure, the total volume of air used to inflate the balloons, $V_T$, will also show hysteresis as a function of the pressure. To describe this behavior, we can use a Preisach model of independent hysterons \cite{preisach1935magnetische}, which we briefly describe here for completeness.
The Preisach model consists of $N$ independent hysterons. Each hysteron can be in one of two distinct states, which we call $0$ and $1$. A hysteron transitions from state $0$ to $1$ when the driving field reaches a threshold value, $H_+$ and from $1$ to $0$ at a threshold value of $H_- < H_+$ (Fig. \ref{fig:hysteron}a) \cite{barker1983magnetic,sethna1993hysteresis,mungan2019structure}. The state of the hysteron between the lower and upper threshold then depends on the history of the driving. When a collection of hysterons have a distribution of lower and upper thresholds, they exhibit return-point memory \cite{shohat2022memory,paulsen2019minimal,middleton1992asymptotic,terzi2020state}.

Because of the hysteresis of traditional, rubber party balloons, the transitions they exhibit at constant pressure are also hysteretic, allowing us to assign a balloon a state, $0$ or $1$, based on the pressure history experienced by that balloon (Fig. \ref{fig:hysteron}b). Note that many balloons also exhibit different curves when inflated and deflated, and their behavior is further complicated by plastic deformations that soften them after repeated cycles of inflation and deflation \cite{johnson1995mullins, de2009damage}. In this paper, however, we will ignore these potential complications and assume our balloons are made of an elastic material with the same inflation and deflation curves. Extending this analysis to balloons with different behavior under inflation and deflation pathways will be explored elsewhere.

We consider a system of $N$ balloons and assume that the $i^{th}$ balloon has transition pressures at $P_{i-} < P_{i+}$ (Fig. \ref{fig:hysteron}b). We choose to label the balloons, with no loss of generality, such that $P_{1+} > P_{2+} > \cdots > P_{N+}$; then it becomes clear that all of the possible transitions between states of $N$ balloons is determined only by the ordering of the local minima, $P_{i-}$.

\begin{figure}[]
  \centering
  \includegraphics[width=8.6cm]{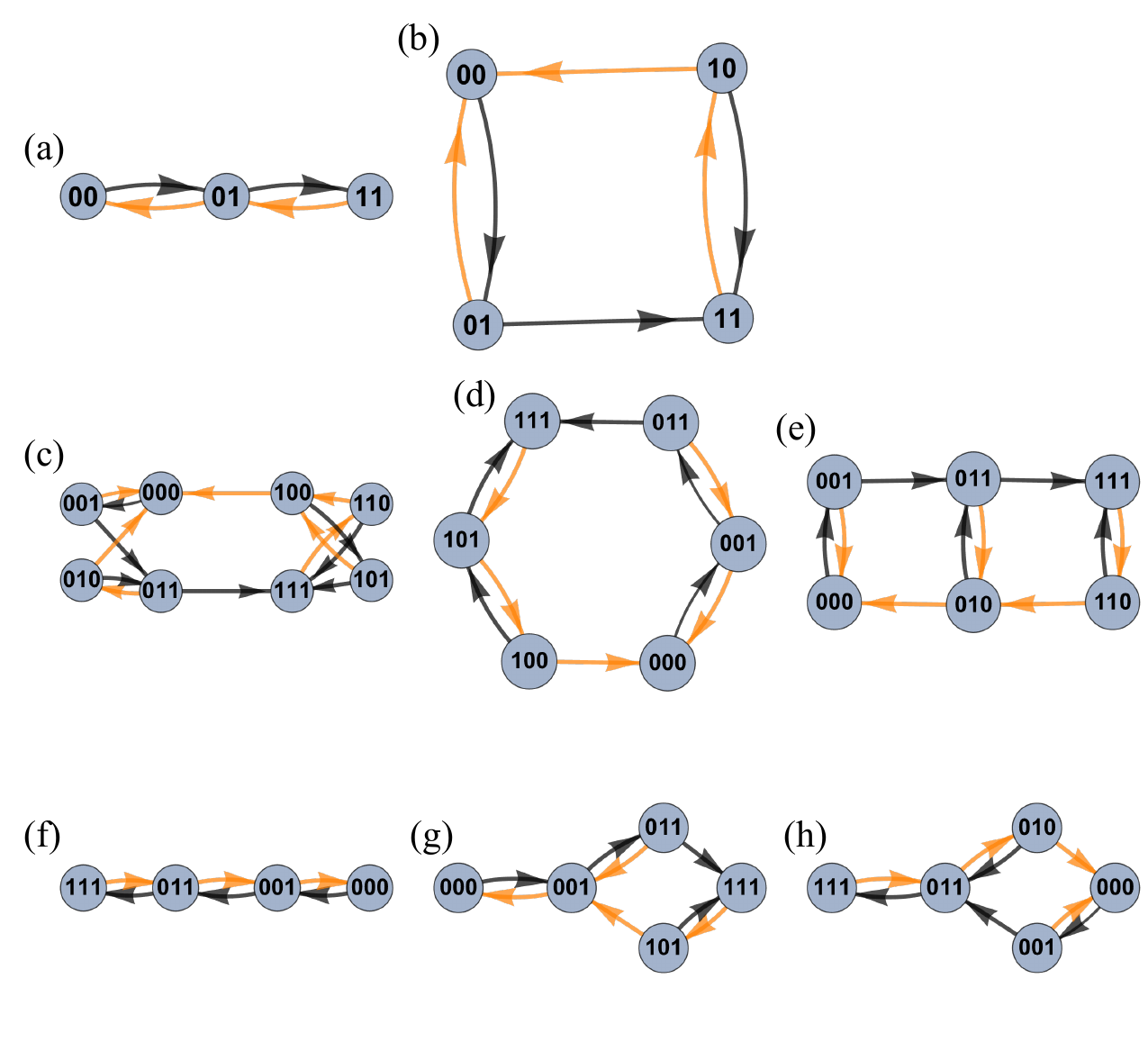}
  \caption{(a-b) Transition graphs for 2 and (c-h) 3 balloons under constant pressure. Black (darker) arrows indicate transitions involving an increasing pressure while orange (lighter) ones are transitions to lower pressure.}
  \label{fig:tgraphs}
\end{figure}

As is standard, we can construct transition graphs as follows: \cite{van2021profusion}
\begin{enumerate}[i)]
    \item Start from the lowest collective state $(0,0, \cdots, 0)$, and determine its ``up'' transition by finding  \[\min_{i}P_{i+}\] This indicates that the $(0,0, \cdots, 0)$ state has a link to the state in which the first balloon to transition has transitioned to the state $1$.

    \item For each new collective state, determine the ``up'' transition by finding \[\min_{i_0}P_{i_{0+}}\] and the ``down'' transition by finding \[\max_{i_1}P_{i_{1-}},\] where the index $i_0$ spans hysterons in state $0$ and $i_1$ spans hysterons in state $1$. These are used to generate new links to the states that can now be reached by the transitions of individual balloons at the given pressures.
   
    \item Stop when all possible transitions have been accounted for.
\end{enumerate} 
For $N$ balloons, there are, in principle, $N!$ possible transition graphs. These are shown explicitly for $2$ and $3$ balloons in Fig. \ref{fig:tgraphs}. It is interesting to notice that though there are $N!$ different transition graphs, the number of graph topologies -- that is, graphs having the same number and \emph{arrangement} of vertices, edges, and loops -- can be fewer.
For three balloons, for example, there are only $5$ topologically distinct transition graphs (Fig. \ref{fig:tgraphs}).

\subsection{\label{sec:volumecontrol} Volume control}

\begin{figure}[]
  \centering
  \includegraphics[width=8.6cm]{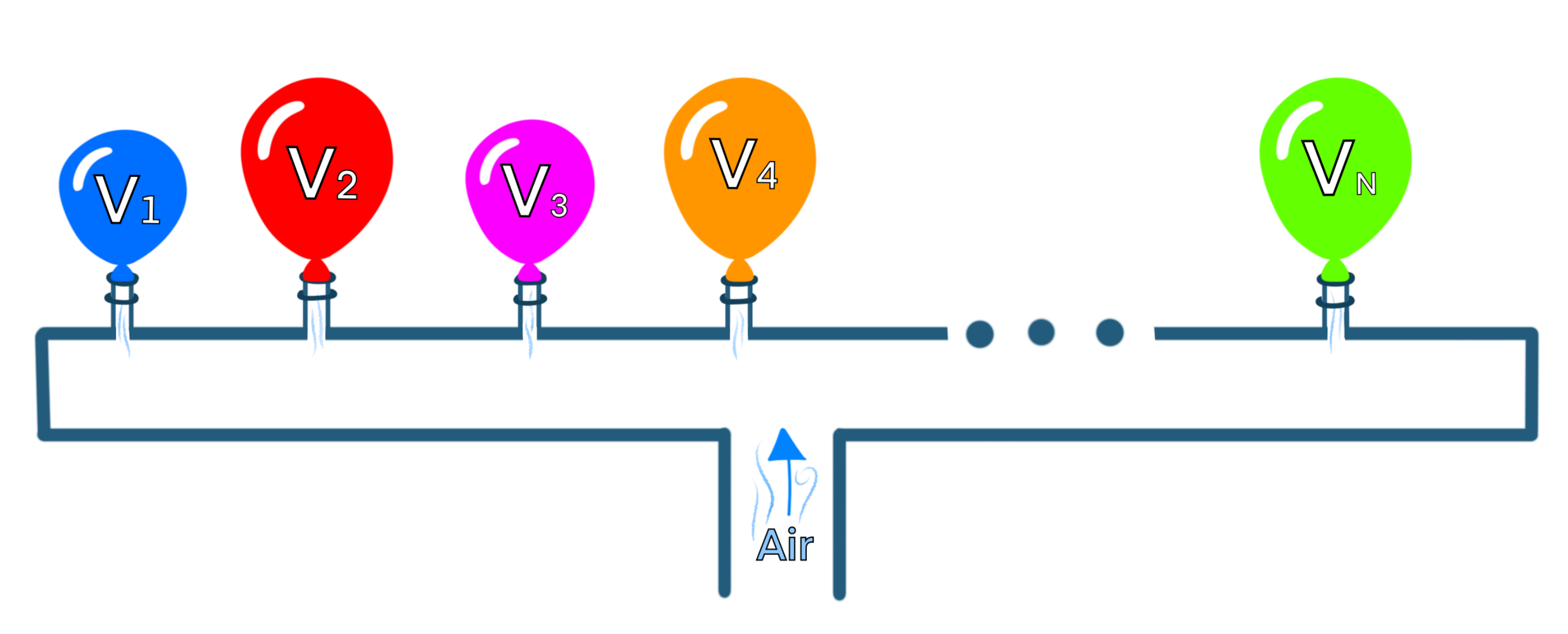}
  \caption{Schematic of the connected system of balloons sharing constant total volume}
  \label{fig:schematic}
\end{figure}

We now turn our attention to a system of $N$ balloons sharing a single reservoir of air (Fig. \ref{fig:schematic}), so that the system satisfies the constraint $\sum_{i=1}^N V_i = V_T$. Rather than controlling the internal pressure, we then consider the behavior of the system as $V_T$ is changed.

If we are to follow the analogy between balloons and magnetic materials, the volume-controlled system is analogous to fixing the magnetization and asking for the magnetic field that produced it. Reflecting on the graph in Fig. \ref{fig:hysteron}a, however, shows that the field associated with a given magnetization in a single hysteron is not even single-valued. However, even though single balloons are hysteretic at constant pressure, their pressure is a function of their volume (Fig. \ref{fig:hysteron}b). This property is shared by many systems of realistic hysterons such as origami bellows \cite{origamiswitches}, elastic conical shells \cite{van2023nonlinear}, and buckling beams \cite{bertoldi2017flexible}.
A critical consequence is that the branch of the pressure-volume curve with negative slope (the dashed curve in Fig. \ref{fig:hysteron}), which is unstable under pressure control, becomes accessible under volume control.

\begin{figure}[]
  \centering
  \includegraphics[width=8.6cm]{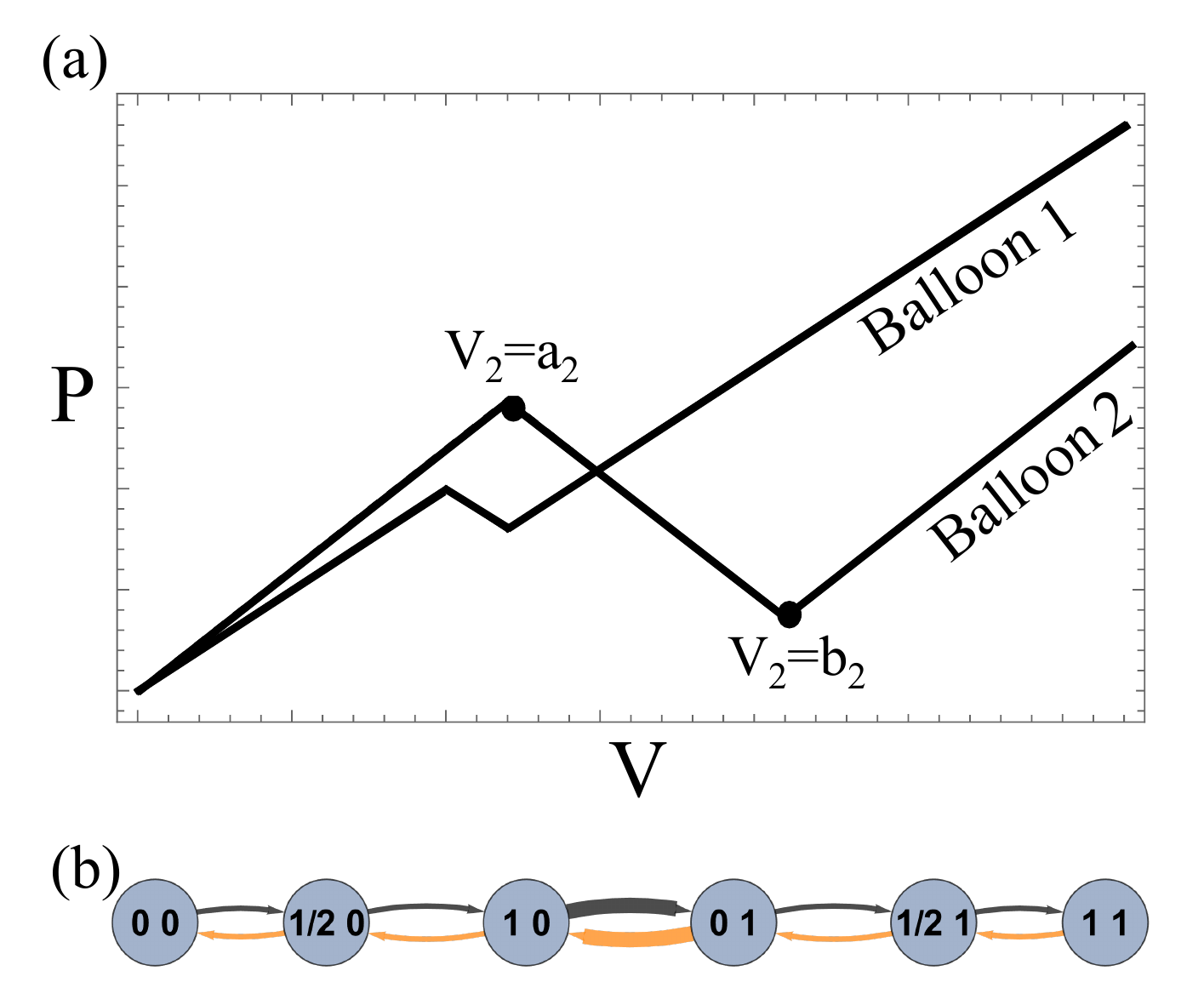}
    \caption{(a) The PV curves of two balloons used to obtain the transition graph under a volume controlled inflation shown in (b), where thicker arrows indicate an avalanche.
    }
    \label{fig:tgraphsavalanche}
 \end{figure}

To simplify the analysis while capturing the essential features, we assume that we can approximate the pressure of a single balloon as a linear function of the volume (Fig. \ref{fig:tgraphsavalanche}a), and keeping the N-shape of the curve. We will show later that this approximation works well for qualitative analysis since the important aspects of the curve are kept. Thus, the pressure of the $j^{th}$ balloon will be approximated by
\begin{equation}\label{eq:linear-pressure}
    P_j = V_j g_j+h_j,
\end{equation}
where
\begin{equation}
    g_j=  \left\{ \begin{array}{lcl}
        -m_j, & &\textrm{if}~s_j = 1/2, \\
        m_j, & & \textrm{otherwise}
    \end{array}\right.
\end{equation}
and
\begin{equation}
  h_j = \left\{ \begin{array}{lcl}
    0, & & \textrm{if} ~ s_j = 0 \\
    2 m_j a_j & & \textrm{if} ~ s_j = 1/2 \\
    2 m_j (a_j - b_j) & & \textrm{if} ~ s_j = 1
  \end{array}\right..
\end{equation}
Here, $a_j$ and $b_j$ are the volumes at which a balloon switches states, $m_j$ is the slope of the pressure as a function of volume, and $s_j$ is the state of the $j^{th}$ balloon. The state $s_j$ determined as follows: it is $0$ if the balloon's volume lies in the first ascending branch of the PV curve, $1/2$ if it lies in the descending branch, and $1$ if the volume lies in the second ascending branch of the curve, as shown in Fig. \ref{fig:tgraphsavalanche}a.
At constant volume, a system of N balloons will equalize their pressure, $P$, at the shared total volume $V_T$. 
\begin{eqnarray}
\label{eq:equalizepressure}
    P_i(V_i) &=& P \\
\label{eq:totalvolumehysteron}
    \sum_{i=1}^N V_i &=& V_T,
\end{eqnarray}
for all balloons. 

Using Eqs. (\ref{eq:linear-pressure}), (\ref{eq:equalizepressure}) and (\ref{eq:totalvolumehysteron}), we can write the total volume as a function of the volume of a single balloon $i$ and the states of the other balloons in the system.
\begin{equation}\label{eq:totalvolume-linear-pressure}
    V_{T}= \sum_{j=1}^n\left(V_i\frac{g_i}{g_j}+\frac{h_i-h_j}{g_j}\right)
\end{equation}
From this, identify the switching fields for each hysteron $i$,
\begin{equation}\label{eq:total-volume-switches}
    V_{T,i\pm}=\sum_{j=1}^n\left(V_{i\pm}\frac{g_i}{g_j}+\frac{h_i-h_j}{g_j}\right)
\end{equation}
where $V_{i+},$ and $V_{i-}$ are the `bare' switching fields, \textit{i.e.} the individual volume values at which the $i^{th}$ hysteron switches its state (either $a_i$ or $b_i$ in Fig. \ref{fig:tgraphsavalanche}, depending on the current state of the balloon). If $s_i=0$, $V_{i+}=a_i$, but if $s_i=1/2$, then $V_{i+}=b_i$. 
The volume, $V_{T,i\pm}$, on the other hand, is the total volume value at which hysteron $i$ changes state. When these switching field values are only dependent on the collective state $\mathbf{s}=(s_1,s_2,...,s_N),$ and not other aspects of the driving history, the system can be described as a system of interacting hysterons \cite{ding2022sequential, kwakernaak2023counting}. To find the up/down transitions from a state $\mathbf{s}$, we need to calculate: \begin{eqnarray}
\label{eq:upstate}
    V_+(\mathbf{s})=\underset{i_0}{\min} V_{T,i_0 +} \\
\label{eq:downstate}
    V_-(\mathbf{s})=\underset{i_1}{\max} V_{T,i_1 -},
\end{eqnarray}
where now $i_0$ runs over hysterons in states $0$ and $1/2$ while $i_1$ runs over hysterons in states $1/2$ and $1.$
We can also rewrite Eq. (\ref{eq:totalvolume-linear-pressure}) to write the individual volumes as a function of the total volume and the collective state of the system, 
\begin{equation}\label{eq:individual-volumes}
    V_i(V_T,\mathbf{s})=\frac{V_{T}-\sum_j\frac{h_i-h_j}{g_j}}{\sum_j g_i/g_j}.
\end{equation}

Under pressure control, the collective state of the system after a transition was always stable, now the stability of an individual balloon depends on the state of the other balloons. The procedure for determining transition graphs is then modified by an additional step. 
For hysterons that do not have an $s=1/2$ state, a detailed account of state stability is given in Ref. \cite{van2021profusion}. In our case, we check for the stability of a landing state as follows. Suppose the system is in a collective state, $\mathbf{s}$, and there is a transition at a total volume $V_T=V_{T_{C\mathbf{S}}}$ that takes it to state $\mathbf{s}'$. The new state is a stable landing spot for the system if the individual volume $V_i(V_{T_{C\mathbf{S}}}+\epsilon,\mathbf{s}')$ for each hysteron $i$ lies within the range of volumes matching its adjective state $s'_i,$ where $\epsilon$ is a very small positive/negative number when the total volume is increasing/decreasing. Cases where a system goes through intermediate unstable states to reach a stable one signify the existence of avalanches as seen in Fig. \ref{fig:tgraphsavalanche} and is a hallmark of interacting hysterons \cite{corrugated,van2021profusion,mungan2019networks}.



A noticeable difference between the interacting system of balloons and the non-interacting one, as seen from the transition graph in Fig. \ref{fig:tgraphsavalanche}, is the occurrence of the state $s=1/2$ in the interacting system of balloons, which arises from the descending branch of the pressure.
This has an obvious effect on what we can observe physically as well. While in the pressure controlled system, there are sudden jumps in the volumes of individual balloons \cite{plants}, the state switching in the volume controlled case does not always involve discontinuous jumps in the  volumes. Instead, transitions from $0$ to $1/2$ or $1/2$ to $1$ are characterized by incremental changes in the volume, except in the case of avalanches. For all the examples in this paper, when there occurs a discontinuous jump in the state of the system, there is only one single stable state it could land in. In general, however, it is impossible to rule out the existence of multiple stable states during avalanches. 

\begin{figure}[]
  \centering
  \includegraphics[width=8.6cm]{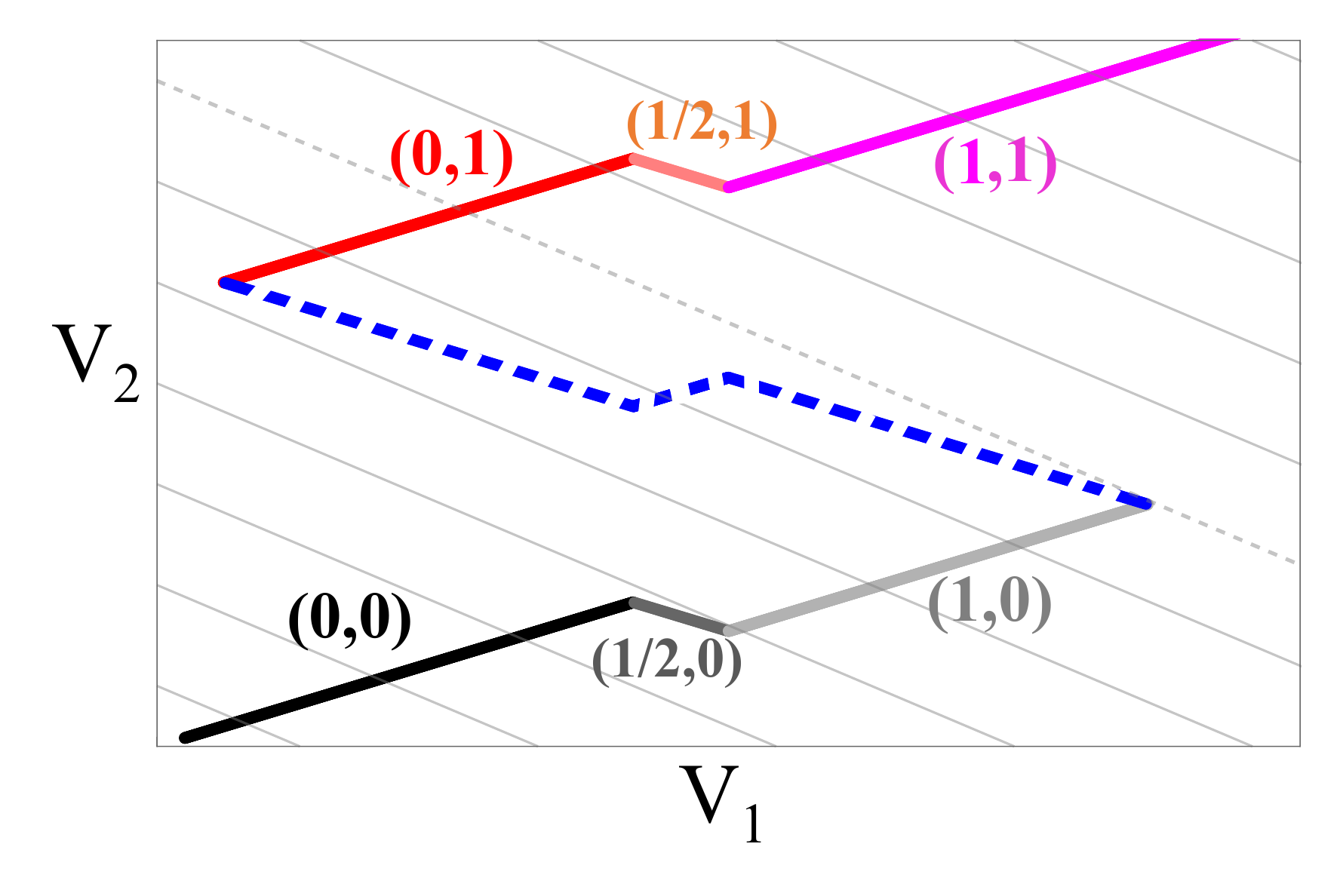}
    \caption{The configuration space for the two balloons from Fig. \ref{fig:tgraphsavalanche}. The dashed gray contour line indicates the constant volume value at which the avalanche occurs, where the system jumps from the state $(1,0)$ to $(0,1)$, while the dashed (dark) lines indicate an unstable solution.
    }
    \label{fig:avalanche}
 \end{figure}

In the next section we demonstrate an alternative pathway to understand the behavior of these systems by tracing the bifurcations of a suitably defined configuration space. Such a configuration space for the linearly approximated system from Fig. \ref{fig:tgraphsavalanche} is shown in Fig. \ref{fig:avalanche}, and details on obtaining it are given in the next section. We believe this might provide a new angle to explore these kinds of systems, as well as uncover new behavior not easily seen with models of interacting hysterons.

\section{The configuration space of $N$ balloons sharing a volume} \label{sec:configspace}
A system of $N$ balloons sharing a constant total volume of air, $V_T$, will equalize their pressure, $P$. We denote the state of the system with an $N+1$ dimensional vector, $\mathbf{V} = (V_1, \cdots, V_N, P)$, satisfying the Eqs. (\ref{eq:equalizepressure}) and (\ref{eq:totalvolumehysteron}). To study the solutions of those equations, we define the vectors
\begin{equation}\label{eq:system2}
    \mathbf{F}(\mathbf{V})=\begin{pmatrix}
        P_1(V_1)-P \\
        \vdots\\
        P_N(V_N)-P\\
        V_1+...+V_N
    \end{pmatrix},\:\:\:\:\:\textrm{and}~
    \mathbf{B} =\begin{pmatrix}
        0 \\
        \vdots\\
        V_T
\end{pmatrix}.
\end{equation}
Then the system of equations takes the compact form, $F_\alpha(\mathbf{V}) = B_\alpha$ where $\alpha$ ranges from $1$ to $N+1$ and indexes the components of $\mathbf{F}$ and $\mathbf{B}$.
Equations of this type occur in a number of contexts, but particularly in bar-joint mechanisms in which bars of fixed length are connected by freely rotating joints.

\begin{figure}[]
  \centering
  \includegraphics[width=8.6cm]{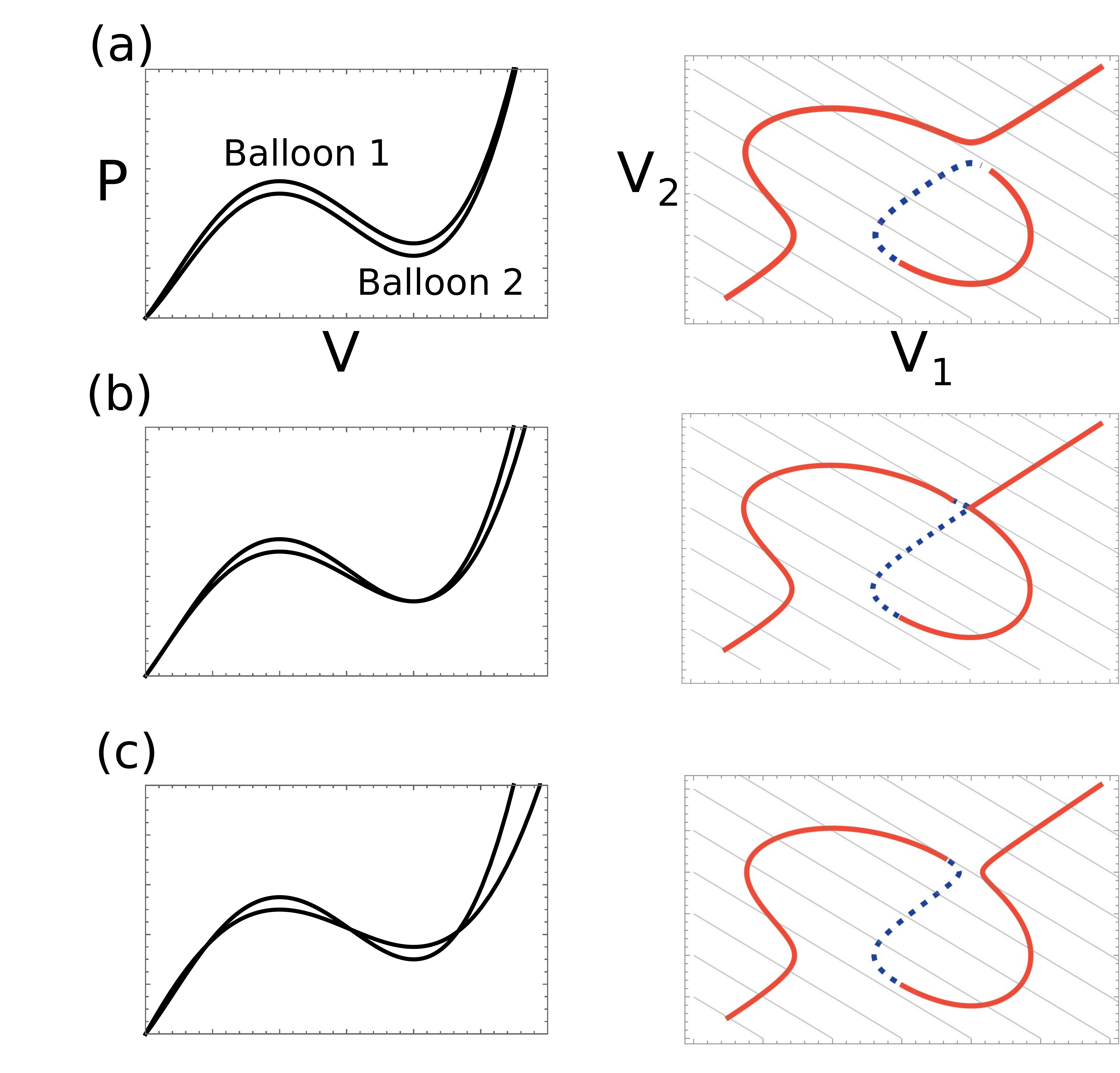} 
  \caption{The configuration space for two balloons inflated under volume control with different pressure versus volume curve combinations. The branch point is shown in (b), where we can see the conic shape of the configuration space at the branch point, and a small deviation from the critical value of the control parameter shows the two hyperbolas formed, in (a) and (c). The diagonal lines shown are contours of constant total volume values. The pressure-volume curves are distinguished through the order of pressure maxima $(P_{1+}>P_{2+}).$ Red (solid) denotes stable solutions while blue (dashed) denotes unstable ones.}
  \label{fig:twoballoons}
\end{figure}

First, whenever the Jacobian of the map $F_\alpha(\mathbf{V})$ is full rank, the system has (possibly many) isolated solutions, $\mathbf{V}(V_T)$. When sweeping over possible values of $V_T$, these solutions trace out curves, as seen in Fig. \ref{fig:twoballoons} with stable solutions drawn as solid curves and unstable solutions drawn as dashed curves. In Fig. \ref{fig:twoballoons}, we have used the smooth pressure from Fig. \ref{fig:hysteron}b. In analogy with mechanisms, we will refer to the space of solutions $\mathbf{V}(V_T)$ as the configuration space of the $N$ balloons. The diagonal light gray lines of constant $V_T$ are provided as a guide to the eye.

\begin{figure}[]
  \centering
  \includegraphics[width=8.6cm]{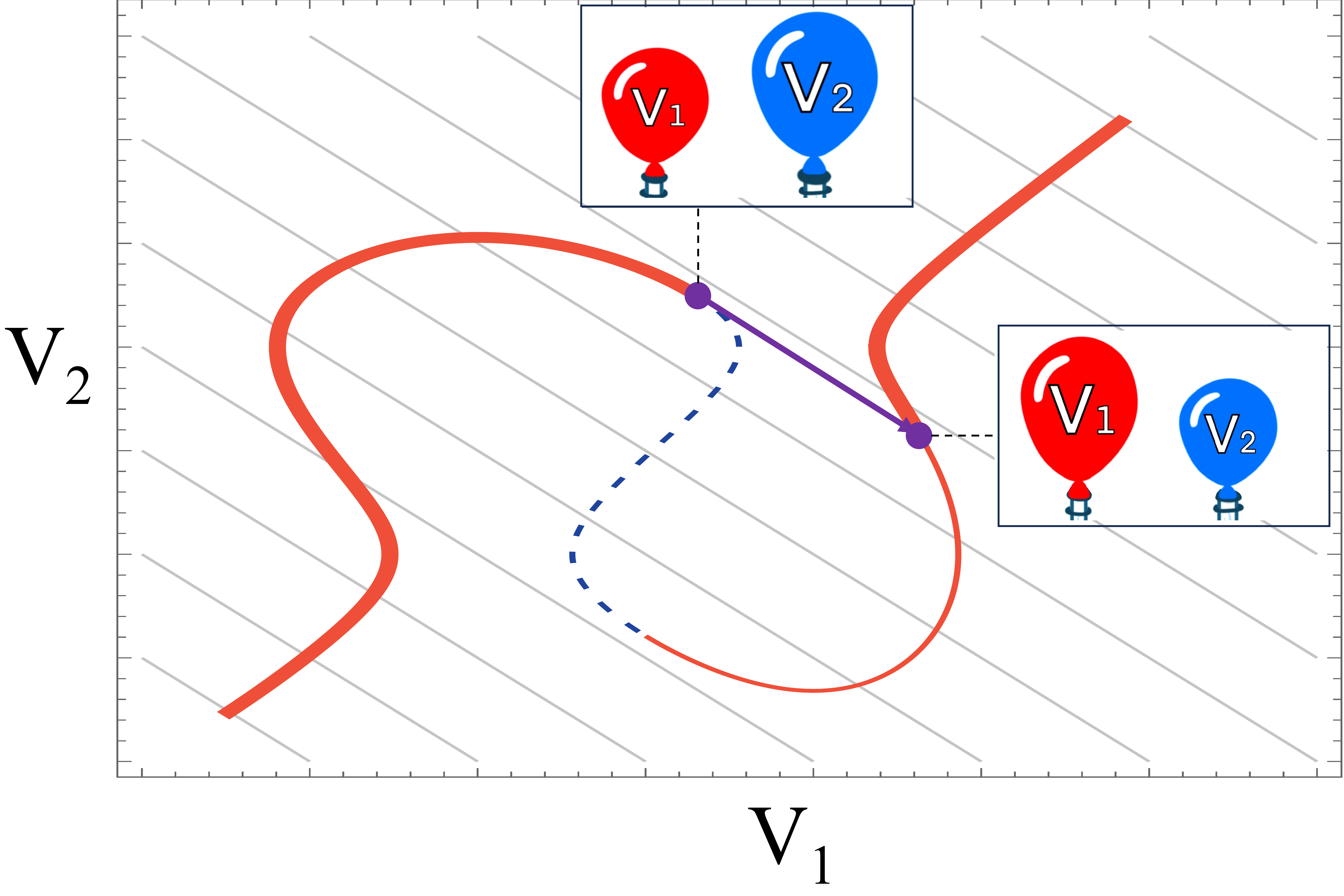} 
  \caption{The configuration space for two balloons inflated under volume control showing the avalanche that occurs at a critical constant volume value. The red (solid) curve indicates a stable solution while blue (dashed) indicates an unstable one. The thicker red curve is the path that the system follows upon inflation. The purple arrow indicates the discontinuous jump that the system undergoes}
  \label{fig:configwithjump}
\end{figure}

Fig. \ref{fig:twoballoons} also makes apparent, however, that there can exist degenerate points for which the Jacobian of $F_\alpha$, $J_{\alpha \beta}(\boldsymbol{V}) = \partial F_\alpha/\partial V_\beta $, is not full rank. For example, a stable and unstable solution can meet at a saddle-node bifurcation, which changes the number of distinct states the system can occupy for a given $V_T$. However, we also see transcritical bifurcations, such as in Fig. \ref{fig:twoballoons}b, which are points where two solutions exchange stability. In the example shown in Fig. \ref{fig:twoballoons}b, such a point has the appearance, and is sometimes called, a branch point. As $V_T$ is tuned through a branch point, the system chooses, essentially randomly, which branch to follow. Notice, however, that the transcritical point in Fig.  \ref{fig:twoballoons}b distinguishes two different ways of connecting the configuration space and, consequently, two different physical behaviors. Fig. \ref{fig:twoballoons}a shows a situation in which the configuration space has two components, but where increasing $V_T$ results in continuous inflation as the system follows the stable branch from one corner of the plot to the other. In \ref{fig:twoballoons}c, however, the configuration space is connected into one component, and now a continuously increasing $V_T$ would lead to a discontinuous jump at a critical volume and, upon deflation, another discontinuous jump at a lower volume. An example of one of these jumps is shown in Fig. \ref{fig:configwithjump}.

Thus, the topology of the configuration space determines the physical behavior, with a branch point separating the two regimes of behavior.
Finally, note that it is the functions $P_i(V_i)$ that determine what kind of physical behavior is seen, with the branch point occurring precisely when the two local minima are at the same pressure. It is a simple consequence of the implicit function theorem that changes in the topology of the configuration space cannot occur except in the presence of a bifurcation. These arguments are outlined in great detail for mechanisms in Ref. \cite{mechanisms} and we do not reproduce them all here.

These observations motivate a search for critical points of $F_\alpha(\mathbf{V})$. To find these, we solve
\begin{equation}\label{eq:Jacobiandet}
    \text{det}(\partial F_\alpha/\partial V_\beta)=\sum_{i=1}^N \left(\prod_{j\neq i}^N P'_j(V_j)\right) = 0.
\end{equation}
and Eq. (\ref{eq:system2}) simultaneously. Both the saddle-node bifurcations for which stable and unstable branches meet as well as the transcritical bifurcation in Fig \ref{fig:twoballoons}b occur as solutions to these equations. By numerically solving Eqs. (\ref{eq:system2}) and (\ref{eq:Jacobiandet}), we have found that the number of bifurcation points, on average, is $10$ for $N=3$ balloons, $45$ for $N=4$, $145$ for $N=5$, $461$ for $N=6$, and $1484$ for $N=7$, which shows a growth of the number of bifurcation points with increasing number of balloons consistent with $\sim e^N$. This growth is consistent with the exponential increase in the number of transition diagrams \cite{van2021profusion}.

Thus, it would be useful to isolate the branch points only. To do this, we treat the total volume, which parametrizes the curves of the configuration space, as a parameter. The state of the system is now described by an augmented vector $\tilde{\mathbf{V}} = (V_1, \cdots, V_N, P, V_T)$; thus, the system of equations now has $N+2$ degrees of freedom but only $N+1$ constraints. We know that there will be zero modes \cite{calladine1978buckminster}, defined as the elements of the right null space of the augmented Jacobian,
\begin{equation}\label{eq:zeromodes}
    J_{\alpha i}(\boldsymbol{V}) \delta V_i =0.
\end{equation}
Now, critical points $\boldsymbol{V}_c$ at which the augmented Jacobian fails to be full rank are characterized by elements of the left null space,
\begin{equation}\label{eq:self-stress}
     \sigma_{\alpha}J_{\alpha i}(\boldsymbol{V}_c)=0.
\end{equation}
In the case of spring networks, they are often called self-stresses \cite{calladine1978buckminster, mechanisms, energetic}, since they represent the collection of spring tensions such that there is net zero force on each of the nodes. However, self-stresses also have a dual meaning as signifier of nonlinearities that lead to kinks or branches in kinematic mechanisms \cite{mechanisms}. In our case, there seems to be no analogous physical meaning of the left null space other than as a signifier of non-smoothness.
The augmented Jacobian can be written as the $(N+1) \times (N+2)$ matrix,
\begin{equation}\label{eq:Jacobian2}
  J=  \begin{pmatrix}
    P_1'(V_1)   &&  0           &&  \cdots  &&  0           &&  0       &&  -1\\
    0           &&  P_2'(V_2)   &&  \cdots  &&  0           &&  0       &&  -1\\
    \vdots      &&  \vdots      &&  \ddots  &&  \vdots      &&  \vdots  &&\vdots\\
    0           &&  0           &&  \cdots  &&  P_N'(V_N)   &&  0       &&-1\\
    1           &&  1           &&  \cdots  &&  1           &&  -1      &&0\\
    \end{pmatrix}.
\end{equation}
It is now explicit that its rank drops by one every time two pressure derivatives are simultaneously zero for some value of $\mathbf{V}$ that is an equilibrium solution of the system. When the maxima and minima of the pressure occur at the same volume in each balloon, this can only happen when $P_{i-} = P_{j-}$ for balloons $i$ and $j$.

We show that the critical points of the augmented Jacobian are, indeed, the branch points observed in the balloon configuration space. We start by expanding $F_{\alpha}(\boldsymbol{V}+\delta \boldsymbol{V})$ for small changes in volumes,
\begin{equation}\label{eq:eqn-expansion}
    J_{\alpha i}\delta V_i+\frac{1}{2} \frac{\partial^2 F_{\alpha}(\boldsymbol{V}_c)}{\partial V_i \partial V_j} \delta V_i \delta V_j + \mathcal{O}(\delta V ^3)=0
\end{equation}
Writing a formal series expansion, $\delta \boldsymbol{V}=\delta \boldsymbol{V}^{(1)}+\delta \boldsymbol{V}^{(2)}+...,$ and substituting this into Eq. (\ref{eq:eqn-expansion}), one finds that $\delta V_i^{(1)}$ is a zero mode of the Jacobian satisfying \cite{tensegrity},
\begin{equation}\label{eq:eqnexpansion2}
    \frac{1}{2}\sigma_{\alpha}^{(n)}\frac{\partial^2F_{\alpha}(\boldsymbol{V}_c)}{\partial V_i \partial V_j}\delta V_i^{(1)} \delta V_j^{(1)}=0,
\end{equation}
where ${\sigma_{\alpha}^{(n)}}$ is a basis for the self stresses at the critical point. 

We assume that all critical points are isolated and that each critical point has only one self stress, which is true as long as no more than two pressure derivatives are simultaneously zero at that critical point. Then, we choose a basis for the zero modes at the critical point, we call that basis $\eta _{n,i}$, and we write $\delta V_i^{(1)}=c_n \eta _{n,i},$ so that,
\begin{equation}\label{eq:Qmn-matrix}
    Q_{nm}c_nc_m=0,
\end{equation}
where the matrix $Q_{nm}$ is given by
\begin{equation}\label{eq:Qmn-matrix-zeromodes}
    Q_{nm}=\eta_{n,i}\eta_{m,j}\sigma_{\alpha}^{(1)}\partial F_{\alpha}(\boldsymbol{V}_c)/\partial V_i \partial V_j.
\end{equation}

Now we can understand the behavior of the system near a critical point of the augmented Jacobian. Under the assumptions that we have made, if the matrix $Q_{nm}$ is either positive or negative definite, we have a ``rigid system.'' That is, there are no non-trivial solutions to the system. It is not clear that this situation can even occur with $N$ balloons. With a combination of mixed positive and negative eigenvalues, however, the geometry of the configuration space at the critical point is that of pair of lines meeting at the branch point and tangent to the solutions of Eq. (\ref{eq:Qmn-matrix}).
We might now ask what happens if we perturb the pressure curves, $P_i(V_i)$, near such a branch point. Clearly this perturbation must separate the branches (since there will no longer be a branch point); this is precisely what we see in Fig. (\ref{fig:twoballoons}). A detailed proof that the branch breaks into one of two hyperbolas can be found in Ref. \cite{mechanisms}.

\begin{figure}[]
  \centering
  \includegraphics[width=8.6cm]{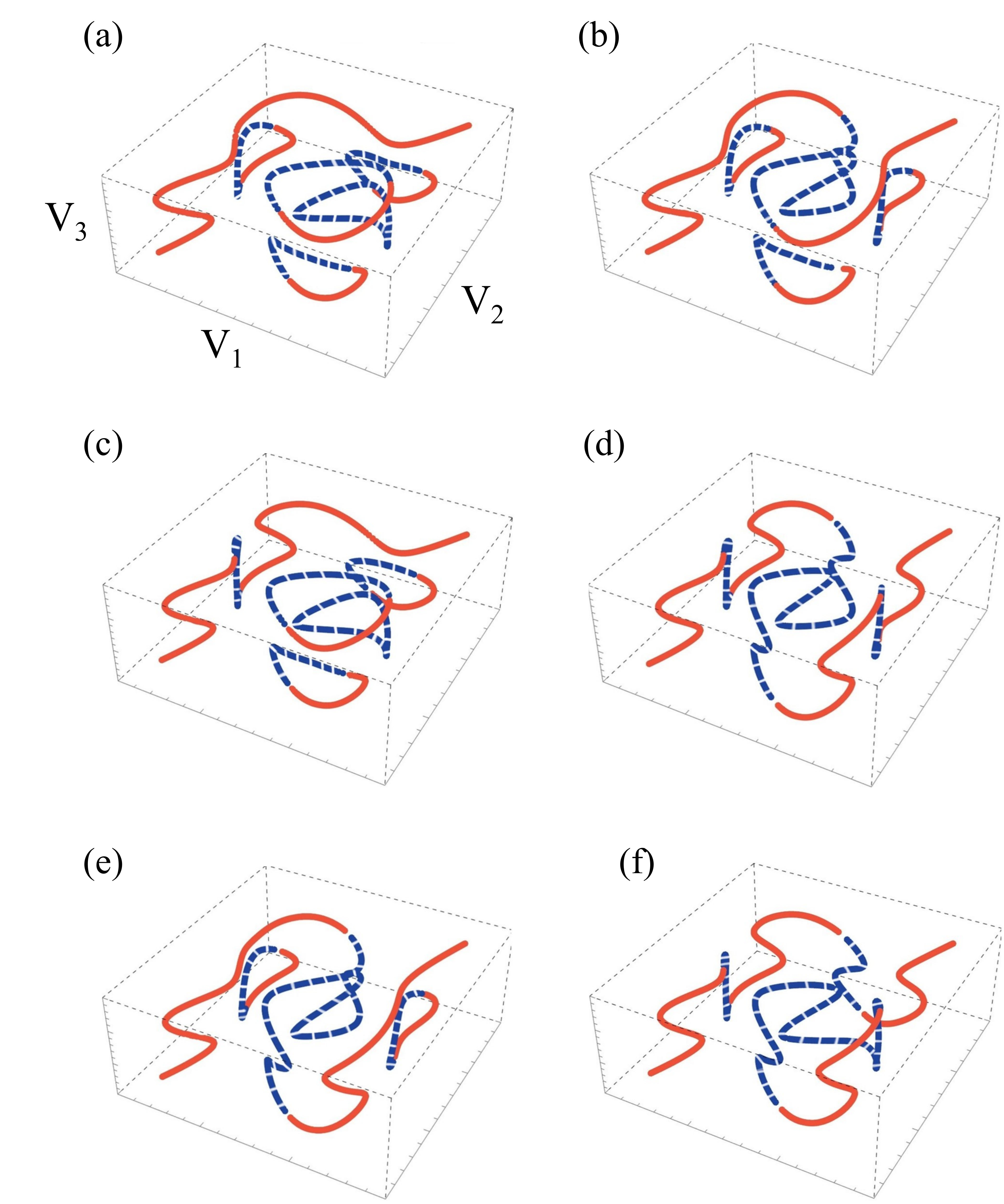} 
  \caption{The configuration space for a three balloons system with various pressure curves minima. Two of the configuration spaces, (b) and (c), have the same topology, namely, they have the same number of components.
  }
  \label{fig:3balloons}
\end{figure}

Finally, in Fig. \ref{fig:3balloons} we show all of the $6$ possible configuration spaces of $3$ balloons. As the number of balloons increases, it appears that the number of individual configuration space components also increases. As the relative pressure curves of the balloons are adjusted, these loops join or separate from each other to create different inflation paths. 
As is the case with two balloons, the topology of these configuration spaces changes through branch points, which occur when two of the balloons pressure derivatives are equal to zero simultaneously for a value $\boldsymbol{V}$ that also solves Eqs. (\ref{eq:equalizepressure}) and (\ref{eq:totalvolumehysteron}). A small perturbation to the pressure curves, once again, results in two hyperbolas forming after the separation of the two meeting branches. That is true, in fact, for a system of any number of balloons as long as only two balloons' pressure derivatives are zero at some value $\boldsymbol{V}$. The analysis is slightly more complicated when more than two are simultaneously zero. 

Taken together, we see that by manipulating the mechanical properties of individual balloons, we can control the order of their inflation or introduce hysteresis. Because of this hysteresis, the history of a single parameter, $V_T$, can be used to determine one of several potential inflation states of the $N$ balloons.

\section{From configuration spaces back to transition graphs}\label{sec:connect}
Finally, we point out that the configuration space approach provides a relatively straightforward route to reconstructing the transition graphs. To begin, it is interesting to note that both the Preisach and the 'volume-control' transition graphs as well as the configuration spaces change topology when the local minima of $P_i(V_i)$ change order. Interestingly, in the case of three balloons, there were only five distinct Preisach transition graphs as two of them were identical by a symmetry; this property is shared by the configuration spaces as well: there are only five topologically distinct configuration spaces in Fig. \ref{fig:3balloons}.

Indeed, it turns out that the transition graphs can be reconstructed from the configuration spaces. To begin, consider two balloons with the simplified piecewise-linear pressure curves in Fig. \ref{fig:tgraphsavalanche}a, for which the configuration space is given by piece-wise linear curves in \ref{fig:avalanche}. Since the states of the transition graphs are labeled by which branch individual balloons reside and these are, consequently, represented entirely by the slope of $P_i(V_i)$, transitions from one set to another are represented by changes in slope in Fig. \ref{fig:avalanche}. In this case, it is straightforward to read off the state of the balloons directly from the slope of the configuration space.

\begin{figure}[]
  \centering
  \includegraphics[width=8.6cm]{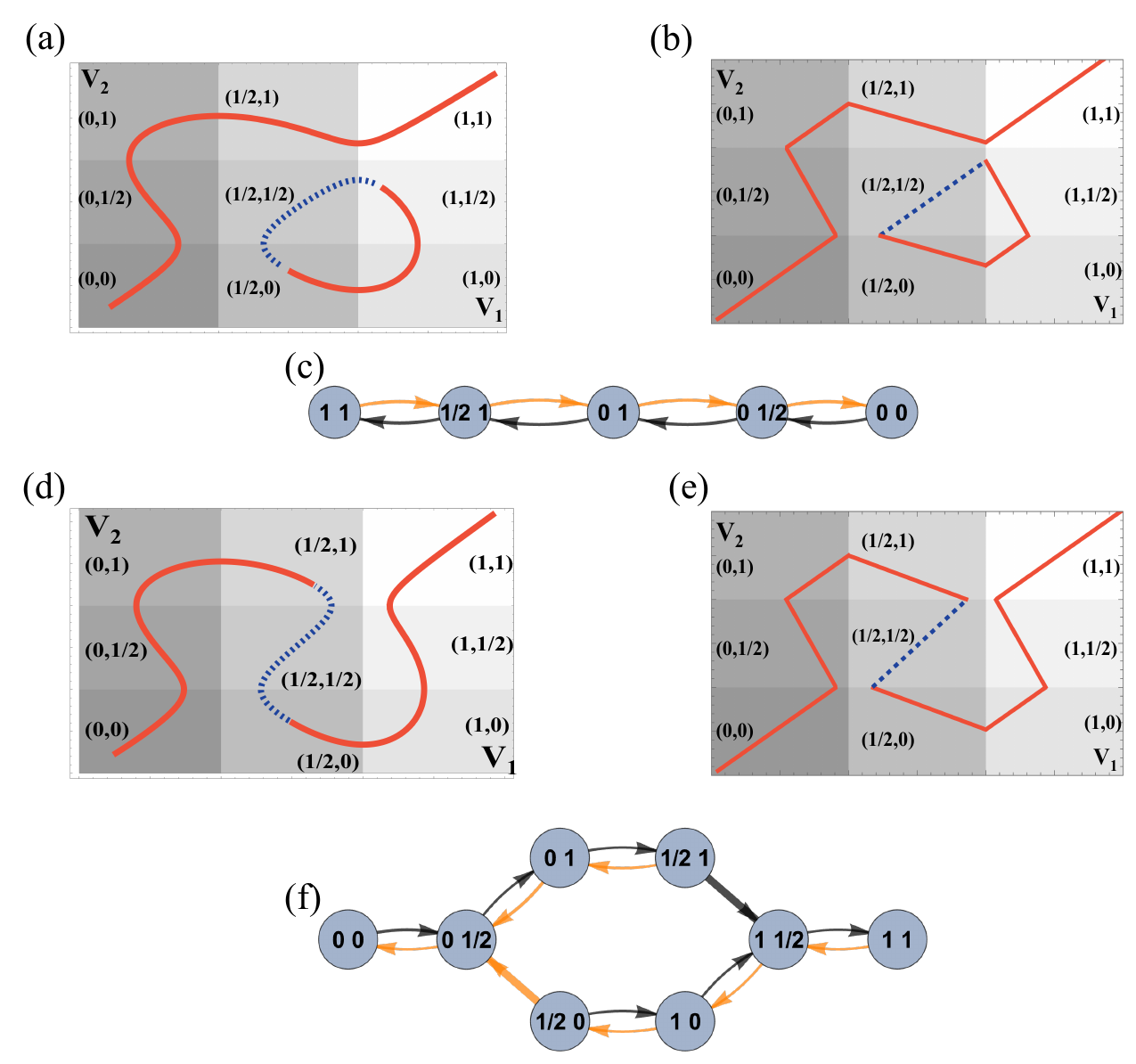} 
  \caption{(a-b) A smooth configuration space and a corresponding linear approximation for two balloons with states labeled in the configuration space. State changes occur when the slopes of the curves change sign. The corresponding transition graph is shown in (c) which follows the main branch, avoiding the `Garden of Eden' states which are, nevertheless, still present in the configuration space. (d-e) A smooth configuration space and corresponding linear approximation for two balloons after a recombination. The transition graph in (f) illustrates the hysteresis that the system exhibits.}
  \label{fig:coloredconfig}
\end{figure}

More generally, the property that $P_i(V_i)$ is a well-defined function means that we can always unambiguously assign a state to the $i^{th}$ balloon only based on its volume relative to $V_{i-}$ and $V_{i+}$. The space $(V_1, \cdots, V_N)$ decomposes into regions, each of which is associated with a unique state of the system in the transition graph. Simple transitions occur when the trajectory of the system leaves one state's domain and enters another. This can be determined whether the trajectories are straight or curved, as seen in Fig. \ref{fig:coloredconfig}. There, we have color coded the regions of the configuration space corresponding to each possible system state to show the relative easiness of constructing the transition graphs off of the space itself. 

Note that when a trajectory changes from a stable to unstable branch, the actual system jumps to a new stable state, and it must do so along the same plane of constant $V_T$. When there is only one other stable state, as in Fig. \ref{fig:avalanche}, it is clear that the system jumps to that new state. As it transitions, it passes along a path that may take it through one or more other states, tracing out the sequence of avalanches. For example, at a certain critical total volume value (the dashed contour line in Fig. \ref{fig:avalanche}), the system must jump from state $(1,0)$ to $(0,1)$.

The configuration space picture also provides more information about how the system finds its ultimate stable state. One can integrate the pressure functions to determine the elastic energy of the balloons and, in particular, the relative energies of all stable states at a specific $V_T$. Though we do not explore it here, this energy landscape also contains information about how the balloons find their ultimate stable state after becoming unstable.

For some values of the total volume, the system can have more than one stable state, though that state may not be accessible from the main branch connected to the deflated state, $\mathbf{V} = (0, \cdots, 0)$. These are sometimes called `Garden of Eden' states, which are states that the system can transition out of but not into them by simply changing the global driving field \cite{origamiswitches}. In the configuration space, they are seen as the loops in Fig. \ref{fig:3balloons} and Fig. \ref{fig:coloredconfig}. In the case of balloons, these metastable states can be accessed by pushing on the larger balloon, forcing the system to jump into the other branch of solutions while keeping the total volume constant.
`Garden of Eden' states should, in fact, be accessible when the system is near a branch point. Moreover, the configuration space can also have loops that are entirely unstable, at least for three or more balloons. Nevertheless, and perhaps surprisingly, the joining and separation of the additional disconnected branches to the main branch plays a crucial role in determining the transition graphs seen in different systems.

\section{\label{sec:conclusion} Conclusion}
As we have seen, systems of balloons can demonstrate complicated and hysteretic behavior, as described by a Priesach model \cite{middleton1992asymptotic, terzi2020state}. When the balloons share a constant volume of air, the hysterons interact through that shared volume, and subsequently allow a richer variety of  transition graphs \cite{van2021profusion}. Because of the descending branch and the fact that a single balloon's pressure is a function of its volume, the transition graphs are more naturally expressed in terms of three states, $0$, $1/2$, and $1$.

Another natural description for this system is, however, as a curve in the configuration space of $N$ balloons described by their respective volumes, $\mathbf{V} = (V_1, \cdots, V_N)$. Different transition graphs of the interacting hysteron model of balloons are represented by these curves leaving and entering distinct regions of the volume space. This gives a geometrical picture of how different transition graphs are related to each other, as well as providing another approach to understanding avalanches. While we focus on small number of balloons for visualization in this paper, the geometrical approach can be applied even for higher numbers of balloons despite the difficulty in visualizing. It would be interesting to explore the statistics of the configuration spaces for large $N$.

One potential complication that would be difficult to fold into the configuration space picture is the fact that some balloons have different inflation and deflation curves. Even if we only increase the total volume, $V_T$, individual balloons can inflate and deflate. Since this effect occurs to individual balloons and the configuration space is a function of individual balloon volumes, this could be incorporated by suitably adjusting the configuration space trajectories according to whether the individual balloons inflate or deflate; this is, in turn, related to the local orientation of the configuration space. This might change where bifurcations happen but should not change the overall picture we lay out.

Many systems that have been traditionally described as interacting hysterons appear to also be in the class of systems of $N$ balloons. In the case of buckling beams \cite{kwakernaak2023counting,bertoldi2017flexible} and bistable origami \cite{origamiswitches}, for example, there is a continuum of strains for which a stress can be assigned, $\sigma(\gamma)$. These systems, therefore, can be analyzed through a similar configuration space lens. Thus, this may be an approach to exploring interacting hysterons in a wide variety of systems.

\begin{acknowledgements}
We are thankful to Samay Hulikal, Sourav Roy, Greg Campbell, Joseph Paulsen, James Pikul, Michael Posa, Ryan Hayward, Mark Yim, and Martin van Hecke for useful comments. The authors are grateful to Yair Shokef for pointing out some errors in a previous version of the manuscript. G.M is thankful to Adea Kabashi for help with illustrations. We acknowledge funding from the National Science Foundation through Grant No. EFRI-1935294.
\end{acknowledgements}

\bibliography{bibliography}%

\end{document}